\documentclass{article}
\input{epsf.tex}
\begin{document}
\newcommand{\HF}{{\scriptsize HF}}
\newcommand{\FM}{{\scriptsize FM}}
\newcommand{\SSDW}{{\scriptsize SSDW}}
\newcommand{\DSDW}{{\scriptsize DSDW}}
\newcommand{\AFM}{{\scriptsize AFM}}
\newcommand{\JPCM}{{\em J. Phys.: Condens. Matter}\ }
\newcommand{\PRL}{{\em Phys. Rev. Lett.}\ }
\newcommand{\PRB}{{\em Phys. Rev. B}\ }
\newcommand{\JPSJ}{{\em J. Phys. Soc. Japan}\ }
\markboth{\small  J. H. Samson, Physics of Magnetism, Pozna\'n 
1996}{\small J. H. Samson, Physics of Magnetism, Pozna\'n 
1996}

\title{PHASE SEPARATION IN THE HUBBARD MODEL}

\author{J. H. SAMSON \\ Department of Physics, Loughborough University, 
\\ Loughborough, Leics LE11 3TU, United Kingdom
\\(Electronic address: j.h.samson@lboro.ac.uk)} 

\maketitle

\begin{abstract}

The Hartree-Fock ground-state phase diagram of the one-dimensional 
Hubbard model is calculated in the $\mu-U$ plane, restricted to phases 
with no charge density modulation.  This allows antiferromagnetism, 
saturated ferromagnetism, spiral spin density waves and a collinear 
structure with unit cell $\uparrow\uparrow\downarrow\downarrow$.  The 
spiral phase is unstable against phase separation near quarter-, half- 
and three-quarter-filling.  For large $U$ this occurs at hole (or 
electron) doping of $(3t/\pi^{2}U)^{1/3}$ from half filling.

PACS numbers 75.10Lp, 75.25+z
\end{abstract}

\section{Introduction}

Although the exact ground state of the one-dimensional one-band 
Hubbard model with nearest-neighbour hopping is well 
known\cite{LiebWu}, systematic Hartree-Fock (\HF) studies of the same model are 
still of value.  They are correct in the limit of large degeneracy, 
they represent an effective Hamiltonian in functional integral 
formalism, and they provide a toy model for understanding the magnetic 
phases of two- and three-dimensional systems (such as cuprate 
superconductors and transition metal alloys).  Such studies also raise 
suggestive connections with state-selection problems in frustrated 
Heisenberg magnets.

The present author has obtained the \HF\ ground state phase diagram of 
the one-dimensional nearest-neighbour Hubbard model\cite{SDW}
\begin{equation}
	H=-t\sum_{i=1}^{N_{\rm 
	a}}\sum_{s=\uparrow}^{\downarrow}\left(c_{is}^{\dag}c_{(i+1)s} + c_{is}^{\dag}c_{(i-1)s}\right)+U\sum_{i=1}^{N_{\rm a}}n_{i\uparrow}n_{i\downarrow}
	\label{Hubbard}
\end{equation}
for arbitrary band filling $n$ ($0\leq n \leq 2$).  The calculation 
was restricted to phases without charge density modulation; subject 
to this restriction, the full $n-U$ phase diagram was obtained.

The work here extends these results to the grand canonical case, 
presenting the $\mu-U$ phase diagram and the asymptotic form of the phase 
boundaries.  Full details of the theory are given in the above 
reference.

\section{Hartree-Fock phase diagram}

The {\it unrestricted} \HF\ approximation minimises the expectation 
value of the Hubbard Hamiltonian (\ref{Hubbard}) in the space of Slater 
determinants.  These states are ground states of a non-interacting 
many-electron system in a potential specified by 
variational parameters.

We restrict consideration to the {\em uniform} phases, where the only 
spatial dependence is in the local magnetisation directions, and to 
macroscopic {\em phase separation}.  In the latter case two uniform phases 
are in equilibrium, separated by a domain wall.  There are two 
families of uniform phases:

\begin{itemize}

 	\item  \SSDW\ (\emph{spiral spin density wave}, 
 	$\uparrow\nearrow\rightarrow\searrow$) of continuously 
 	varying wave vector $Q$.  This has limiting cases
 	\begin{description}
		\item  $Q=0$: \FM\ (\emph{saturated ferromagnetism}, 
		 $\uparrow\uparrow\uparrow\uparrow$).
		 
		 \item  $Q=\pi$: \AFM\ (\emph{antiferromagnetism}, 
		 $\uparrow\downarrow\uparrow\downarrow$).
	 \end{description}
 	\item  \DSDW\ (\emph{double spin density wave}, or two interpenetrating 
 	antiferromagnetic sublattices with N\'eel vectors canted at 
 	an angle $\theta$, 
 	$\uparrow\nearrow\downarrow\swarrow$).
 
  \end{itemize} 

For each point $(n,U)$ the energy of each family is minimised with 
respect to the exchange splitting and the angle ($Q$ or $\theta$).  
The \HF\ energy $E_{\rm HF}(n,U)$ is the lower of the minima for 
\SSDW\ and \DSDW.

\begin{figure}[h]
	\epsfxsize=11cm
	\epsfbox{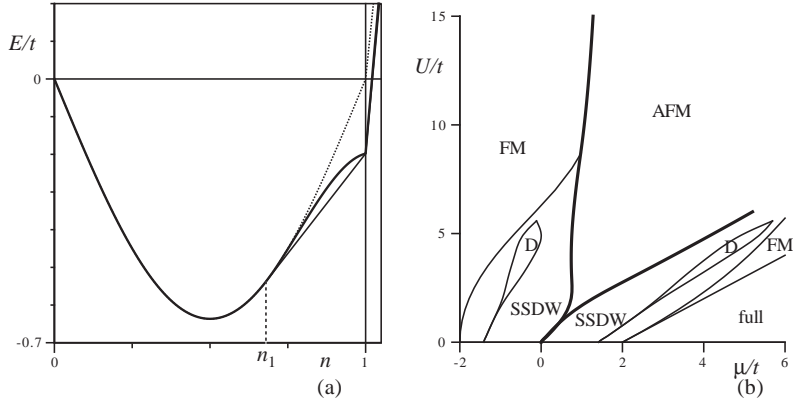}
	\caption{(a) The \HF\ energies for $U=10t$.  Bold curve: \SSDW\ 
	energy; dotted curve: \FM\ energy; thin line: energy of 
	phase-separated state.  (b) The \HF\ phase diagram of the Hubbard 
	model.  The \AFM\ phase has filling $n=1$, and the collinear \DSDW\ phases 
	(denoted `D') have filling $n=1/2, 3/2$.  The region marked `full' 
	corresponds to filled bands ($n=2$).}
	\protect\label{fig}
\end{figure}

The condition for stability 
against macroscopic phase separation is that $E_{\rm HF}(n,U)$ be a convex 
function of $n$, or equivalently that the chemical potential
\begin{equation}
	\mu = \partial E_{\rm HF}/\partial n
	\label{mu}
\end{equation}
be an increasing function of $n$.  If this does not hold, a Maxwell 
construction\cite{MPP}, shown in figure 
\ref{fig}(a), determines the fraction of each phase.

The figure shows the generic picture for $U>8.7t$.  For $n_{1}<n<1$ 
(and $1<n<2-n_{1}$) the \HF\ ground state consists of a uniform \AFM\ 
phase with $n=1$ in equilibrium with a hole-rich (electron-rich) 
uniform \FM\ phase with $n=n_{1}$ ($n=2-n_{1}$).  The chemical 
potential is pinned within the \AFM\ gap and within the lower (upper) 
band of the \FM\ phase.  Additional carriers will simply move the 
domain wall.  A similar effect is seen in supercell calculations of 
collinear configurations\cite{AES}.  For $n<n_{1}$ (and $2-n_{1}<n$) 
the ground state is \FM. Thus the \SSDW\ is always unstable against 
separation into \AFM\ and \FM\ phases.  (Note that the \SSDW\ energy 
peels off from the \FM\ energy, indicating $Q>0$, only above $n_{1}$.) 
The situation is slightly different for $U<8.7t$: near $n=1$ the 
\SSDW\ is unstable against phase separation into an \AFM\ phase and a 
\SSDW\ of longer wavelength.

The previous paper\cite{SDW} presented the phase diagram in the 
$n-U$ plane.  Figure \ref{fig}(b) shows the 
phase diagram in the $\mu-U$ plane.  The areas shown here are pure 
phases, separated by second-order transitions (between \FM\ and 
finite-$Q$ \SSDW) and first-order transitions (between \SSDW\ and 
collinear \DSDW, between \SSDW\ and \AFM, and between \FM\ and 
\AFM).  For large $U$, as a function of chemical potential there are two 
discontinuities in filling factor and in wave number.  As $U$ is 
decreased, further discontinuities appear (although only the \DSDW\ 
phase has been calculated).

\section{Form of phase boundaries}
We can compute the asymptotic form of the \FM--\AFM\ phase boundary for large $U$.  
By expanding the \HF\ energy of the \AFM\ state at $n=1$ we find the 
gap to be
\begin{equation}
	\Delta = U - 4t^{2}/U + 4t^{4}/U^{3} + \cdots
	\label{gap}
\end{equation}
and the energy
\begin{equation}
	E_{\rm HF}(1,U) = - 2t^{2}/U + 2t^{4}/U^{3} + \cdots.
	\label{EAFM}
\end{equation}
(This of course exceeds the Bethe {\it Ansatz} ground state energy $-4\ln 2 
t^{2}/U$\cite{Fulde}.)
Drawing a tangent to the \FM\ energy $-(2t/\pi)\sin n\pi$ gives the 
form of the phase boundary for large $U$ as 
\begin{equation}
	n_{1} = 1 - \left(\frac{3}{\pi^{2}}\frac{t}{U}\right)^{1/3} - 
	\frac{t}{10U} + \cdots
	\label{nps}
\end{equation}
or
\begin{equation}
	\mu = 2t - \left(\frac{3\pi t}{U}\right)^{2/3}t + \cdots.
	\label{mups}
\end{equation}

\section{Discussion}

The full \HF\ solution to the one-dimensional Hubbard model is more 
difficult to compute, in general requiring a $4N_{\rm a}$-dimensional 
minimisation.  One might speculate on a possible devil's staircase for 
small $U$: tongues, such as that for \DSDW, would exist for all 
rational fillings.  Similar behaviour is seen in the Falicov-Kimball 
model (where one spin state is immobile)\cite{FK}.  For larger $U$ (in 
the Hubbard model) the tongues would disappear, thinning out the 
staircase.  We would indeed expect the uniform \SSDW\ phases to 
distort in such a way as to open a gap at the Fermi surface; the 
\DSDW\ can be seen in such a way\cite{ST}.  The phase separation seen 
here may appear in a microscopic form (as a soliton lattice) rather 
than the macroscopic form discussed here.  Realistic terms, absent in 
the one-band Hubbard model, that suppress long-wavelength charge 
fluctuations would tend to prevent macroscopic phase separation.

It is also interesting to note connections with the problem of state 
selection in frustrated Heisenberg magnets.  \DSDW\ phases with 
varying $\theta$ are strictly degenerate in a classical Heisenberg 
model; here it is the itinerant nature of the magnetism that selects 
the collinear state.

We end by confessing that these results bear little relation 
to the true ground state of the one-dimensional one-band Hubbard 
model, which does not break symmetry in this way.  However, they may 
provide a useful starting point for the Hubbard model in two and three 
dimensions and many-band Hubbard models in one dimension.


\begin{thebibliography}{9}
	\bibitem{LiebWu}  E. H. Lieb, F. Y. Wu \PRL\ {\bf 20}, 1445--8 
	(1968).

	\bibitem{SDW}  J. H. Samson \JPCM\ {\bf 8}, 569--80 (1996).

	\bibitem{MPP}  M. Marder, N. Papanicolaou, G. C. Psatalkis, \PRB\ 
	{\bf 41}, 6920--32 (1990).

	\bibitem{AES} A. N. Andriotis, E. N. Economou, C. M. Soukoulis, 
	\JPCM {\bf 5}, 4505--18 (1993).
	
	\bibitem{Fulde}  See e.g. P. Fulde, {\em Electron Correlations in 
	Metals and Solids}, p 290 (Springer, Berlin, 1991).
	
	\bibitem{FK}  J. Lach, R. \L y\.{z}wa, J. J\c{e}drzejewski, \PRB\ {\bf 
	48}, 10783--7 (1993).
	
	\bibitem{ST} Y. Suzumura, N. Tanemura, \JPSJ {\bf 64}, 2298--301 
	(1995); N. Tanemura, Y. Suzumura, \JPSJ {\bf 66} in press (1996).
	
\end{thebibliography}
\end{document}